\documentclass[twocolumn, amsmath, amssymb, aps, superscriptaddress]{revtex4-2}
\usepackage[dvipsnames]{xcolor}
\usepackage{graphicx}
\usepackage{enumitem}
\usepackage{siunitx}
\DeclareSIUnit\gauss{G}
\usepackage{booktabs}
\usepackage{multirow}
\usepackage{mathtools}
\usepackage{flushend}
\usepackage[switch]{lineno}
\usepackage[utf8]{inputenc}
\usepackage{soul,xcolor}
  \makeatletter
    \renewcommand\@make@capt@title[2]{%
     \@ifx@empty\float@link{\@firstofone}{\expandafter\href\expandafter{\float@link}}%
      {\textbf{#1}}\@caption@fignum@sep#2\quad}%
    \makeatother
   
\makeatletter 
\renewcommand{\fnum@figure}{\textbf{Fig.~\thefigure}}
\makeatother
\usepackage{braket}

\usepackage{times}
\usepackage{bm}
\usepackage{amsmath}
\usepackage{placeins}

\newcommand{\probP}{\text{I\kern-0.15em P}}

\newcounter{lastnote}

\begin{document}

\title{Demonstration of a programmable optical lattice atom interferometer}

\author{Catie LeDesma}
\affiliation{JILA \& Department of Physics, University of Colorado Boulder, Colorado 80309, USA}

\author{Kendall Mehling}
\affiliation{JILA \& Department of Physics, University of Colorado Boulder, Colorado 80309, USA}

\author{Jieqiu Shao}
\affiliation{College of Engineering and Applied Science, University of Colorado Boulder, Boulder, Colorado 80309, USA}

\author{John Drew Wilson}
\affiliation{JILA \& Department of Physics, University of Colorado Boulder, Colorado 80309, USA}

\author{Penina Axelrad}
\affiliation{Department of Aerospace Engineering, University of Colorado Boulder, Boulder, Colorado, USA}

\author{Marco Nicotra}
\affiliation{College of Engineering and Applied Science, University of Colorado Boulder, Boulder, Colorado 80309, USA}

\author{Dana Z. Anderson}
\affiliation{JILA \& Department of Physics, University of Colorado Boulder, Colorado 80309, USA}
\affiliation{Infleqtion, 3030 Sterling Circle, Boulder, Colorado 80301, USA}

\author{Murray Holland}
\email[Correspondence should be addressed to ]{murray.holland@colorado.edu}
\affiliation{JILA \& Department of Physics, University of Colorado Boulder, Colorado 80309, USA}

\date{\today}

\begin{abstract}

\noindent Performing interferometry in an optical lattice formed by standing waves of light offers potential advantages over its free-space equivalents since the atoms can be confined and manipulated by the optical potential. We demonstrate such an interferometer in a one-dimensional lattice and show the ability to control the atoms by imaging and reconstructing the wavefunction at many stages during its cycle. An acceleration signal is applied and the resulting performance is seen to be close to the optimum possible for the time-space area enclosed according to quantum theory. Our methodology of machine design enables the sensor to be reconfigurable on the fly, and when scaled up, offers the potential to make state-of-the art inertial and gravitational sensors that will have a wide range of potential applications.

%\newline\vspace*{1pc}

%Atom interferometry coherently splits the trajectory of atomic wavepackets into multiple pathways and recombines them to produce quantum interference. In this paper, we experimentally demonstrate atom interferometry in a shaken optical lattice, where atoms are subject to a standing wave of light formed by counterpropagating lasers. Performing interferometry in a lattice offers potential advantages over its free-space equivalents since the atoms can be confined, controlled, and manipulated by the optical potential. This offers the possibility to make the sensor reconfigurable on the fly, e.g., into an accelerometer, gyroscope, or gravity gradiometer, as desired. It also allows the atoms to be held for long periods, and to bring the instrument out of the laboratory and operate it in adverse environments where it may be subject to size constraints or to thermal or vibrational noise. We demonstrate such an interferometer in a one dimensional lattice and show the ability to control the atoms by imaging and reconstructing the wavefunction at many stages during its cycle. An acceleration signal is applied and the resulting performance is seen to be close to the optimum possible for the time-space area enclosed according to quantum theory. When scaled up, this approach offers the potential to make state-of-the art inertial and gravitational sensors that will have a wide range of potential applications.

\end{abstract}

\maketitle 
\subsection*{Introduction}
\begin{figure*}[!htb]
    \includegraphics[width= .9\textwidth]{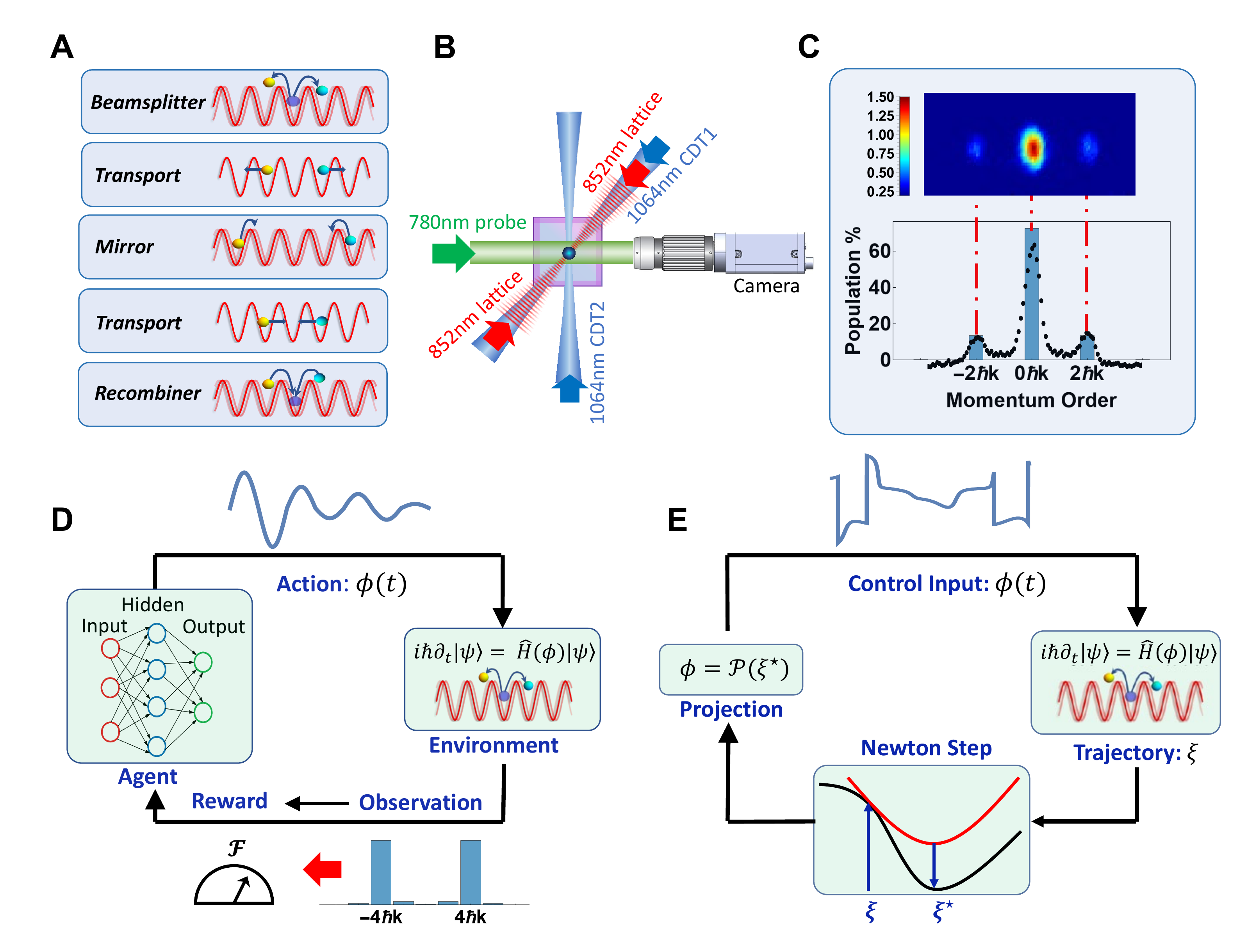}
    \caption{Experimental system: (A) The lattice is shaken to implement the interferometer components. (B) Condensates are created in two crossed 1064~nm dipole laser beams (blue) that intersect inside a science cell (purple). The atoms are then loaded into a one-dimensional 852~nm optical lattice (red), whose position can be altered by applying frequency shifts to two acousto-optic modulators (AOMs; not shown). After time of flight (TOF), an absorption image of the cloud is taken by a 780~nm probe laser (green). (C) A TOF absorption image of the optical density (OD) of the cloud after lattice load. Also shown is the integrated momentum distribution (lower) overlaid with the theoretical histogram anticipated for the lowest Bloch eigenstate at that lattice depth. Machine-design methods: (D) Reinforcement learning, where an agent (neural network) chooses an action, the environment responds to the action, and a reward is given as feedback based on the observed state. (E)~Quantum optimal control using the PRojection Operator Newton method for Trajectory Optimization (PRONTO) algorithm~\cite{PhysRevA.105.032605}. For each iteration, given the control input $\phi(t)$, PRONTO first computes the trajectory~$\xi$ based on the Schr\"odinger equation, then the algorithm uses the second order approximation of the cost, which is a function of $\xi$, to find an updated trajectory~$\xi^{\star}$.}
    \label{fig:schematic}
\end{figure*}

%\subsubsection*{Introduction}
In this paper, we utilize machine learning and optimal control methods to manipulate atoms confined in an optical lattice to sense accelerations. Atom interferometry is well-established as a means of measuring inertial forces with exquisite sensitivity~\cite{PhysRevLett.125.191101, Lee2022}.  Scientifically compelling endeavors that can be advanced through precision interferometry often present constraints or environments that are challenging for typical atom interferometric systems to accommodate. These include low orbit monitoring of the gravity field of Earth~\cite{Tino2019,Lachmann2021,Trimeche_2019}, and spaceborne searches for dark matter~\cite{doi:10.1126/science.aaa8883,Tsai2023}, both of which are constrained by size and weight limitations and present  harsh vibrational and thermal environments.  In another precision measurement context, timekeeping systems that confine atoms in optical lattices have achieved stunning levels of precision---displaying a fractional frequency uncertainty of $3.5\times10^{-19}$~\cite{kim2023} for shallow lattices, and $7.6\times10^{-21}$~\cite{Bothwell2022} for deep lattices. Timekeeping experiments have established that an optical lattice can provide a pristine environment for precision metrology~\cite{Katori2011}, even while the forces imposed on atoms correspond to the range of tens to hundreds of $g$'s, where $g$ is the acceleration due to the gravity of Earth. From a practical standpoint, therefore, optical lattices can be used to confine and manipulate atoms in the face of a dynamically harsh environment~\cite{Bongs2019}; the question then arises whether this system can be used for matter-wave interferometry. The answer lies in machine-learning methods, which can be used to discover how an atomic wavefunction associated with a lattice can be manipulated by changing the relative phase of the optical lattice to achieve interferometric measurement precision~\cite{PhysRevResearch.3.033279,chih2022train}, see Fig.~\ref{fig:schematic}(A).

Optical lattices have been used to enhance the performance of atom interferometric inertial sensors in a variety of ways, such as to demonstrate the high sensitivity of a sensor based on Bloch oscillations~\cite{PhysRevA.102.053312}, to capitalize on long atomic coherence times by holding separated atoms in place~\cite{Panda2024}, and as a means to impose large momentum transfer to atoms  ~\cite{Nelson.2020, Gebbe2021}.  
In this paper, we demonstrate an interferometry sequence where the atoms are confined within the optical lattice potential during the entirety of the measurement sequence. Unlike other optical lattice based sensors, our device efficiently transfers atoms into the conduction band of the optical periodic potential, enabling the atoms to enclose an extended enclosed area for enhanced sensitivity to inertial signals. We call this method Bloch-band interferometry (BBI) due to the fact that during the free propagation of our sensing sequence, the lattice remains stationary and the atomic wavefunction can be described as being in a superposition of Bloch states. BBI combines the approach of the shaken-lattice interferometer~\cite{PhysRevA.64.063613, PhysRevA.95.043624}, where the lattice is dynamically shaken to enable atoms to move through the valence band of the lattice via dynamic tunneling~\cite{PhysRevLett.99.220403}, and that of lattice-based interferometry where atoms are held stationary in a lattice during phase accumulation. In BBI, the lattice is only shaken during application of interferometery protocols, i.e. beamsplitters and mirrors, and is held stationary during propagation steps, allowing atoms to traverse the lattice as essentially free particles rather than through tunneling. The power of this optical lattice approach combined with optimal control methods to generate a nearly arbitrary quantum state was illustrated in Ref.~\cite{PRXQuantum.2.040303}. Motivated by these earlier works here we demonstrate a machine-designed optical lattice atom interferometer and evaluate its performance as a sensor.

\vspace*{-1pc}
\subsection*{Experimental Procedure}
\vspace*{-.75pc}
\begin{figure*}[!ht]
    \centering
    \includegraphics[width=\textwidth ]{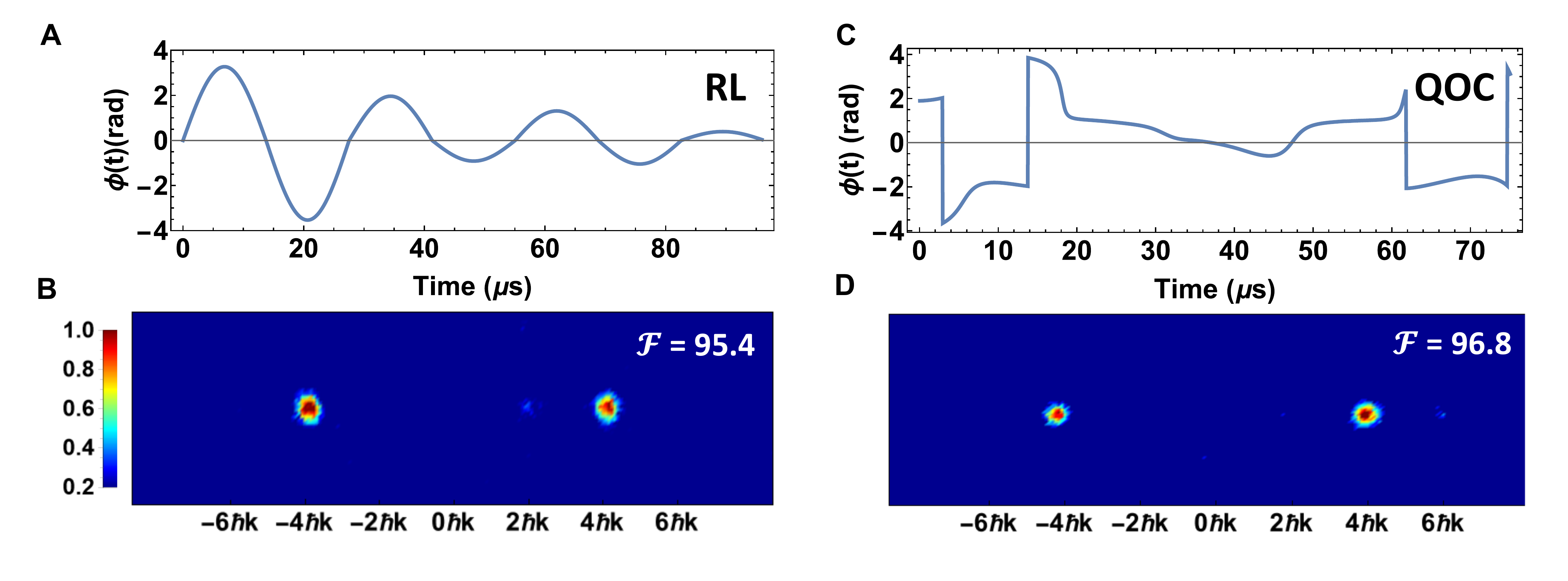}
    \caption{Beamsplitter sequences (A) and (B): Sequence learned by reinforcement learning (RL) with the resulting time of flight images from experiment (A) and (B). Similar measured momentum distributions for quantum optimal control (QOC) are depicted in (C) and (D). The experimental fidelities for these sequences to the target solutions-$\cal F$, as defined in Ref.~\cite{PRXQuantum.2.040303}, are as indicated. In (B) and (D) color bars denote the optical density.}
    \label{fig:expt_bs}
\end{figure*}

%Our experiments begin by loading a pure BEC consisting of approximately $2\times 10^4$ $^{87}$Rb atoms into an optical lattice. 
As shown in Fig.~\ref{fig:schematic}(B), we produce $2\times 10^4$ Bose-condensed $^{87}$Rb atoms via all-optical evaporation of atoms in a 1064~nm crossed dipole trap (CDT)~\cite{PhysRevLett.87.010404}. %We can produce condensates consisting of as many as $10^5$ atoms in about ten seconds, but intentionally reduce the atom number by lowering the depth of the dipole trap at the end of evaporation to minimize the impact of atom interactions which are not currently considered in our learning methods. Since optical evaporation does not require the atoms to be in a particular hyperfine state, our atoms are not spin-polarized.
At the end of evaporation, the dipole beams are maintained at their final setpoint to offset gravity while the optical lattice beams are adiabatically ramped on, leading to the transfer of condensed atoms into the ground-state wavefunction with the momentum distribution shown in Fig.~\ref{fig:schematic}(C). The lattice is formed by two counterpropagating 852~nm laser beams, whose intensity and phase are each controlled by independent acousto-optic modulators (AOMs) before entering the science cell. The atoms are subject to an optical potential at point $\bm{r}$ and time $t$:
\begin{equation}
    V(\bm{r},t) = V_L(\bm{r}_{\perp})\cos\bigl(2\bm{k}\cdot\bm{r} + \phi(t) \bigr) + V_D(\bm{r})
\end{equation}
where wavevector $\bm{k}$ points along the lattice direction (with magnitude $k=2\pi/\lambda$ given by the wavelength $\lambda$ of the light), and $V_L(\bm{r}_{\perp})$ is the depth of the lattice with $\bm{k}\cdot\bm{r}_{\perp}=0$. We refer to $\phi(t)$ as a control, or shaking function. The potential from the CDT, $V_D(\bm{r})$, which remains present during the shaking, gives rise to an ellipsoidal trap~\cite{Liao:17} with frequencies of 36.5, 132.3, and 144.8 Hz, defined along the principal axes and measured via parametric heating~\cite{PhysRevLett.31.1279}. Total atom numbers are calibrated by means of time of flight (TOF) expansion and fitting to the Thomas-Fermi scaling ansatz~\cite{PhysRevLett.77.5315, Dalfovo_1997}. To measure the momentum distribution, all confining beams are rapidly extinguished and the atoms are allowed to expand via TOF for 15~ms. Normalized atom numbers for each momentum state are extracted from the integrated optical density of the probe image. To accurately calibrate our lattice depth, we either perform Kapitza-Dirac diffraction~\cite{PhysRevLett.56.827,PhysRevLett.79.784} and fit with theory or measure the frequency of momentum state oscillations following a small sudden shift of the optical lattice position~\cite{PhysRevA.97.043617}.

\vspace*{-1pc}
\subsubsection*{Reinforcement learning and optimal control methods}
\vspace*{-0.75pc}

The problem of quantum design has been recast in recent years by the developments in machine-learning methods, which can be extremely effective at finding efficient strategies for accomplishing complex tasks \cite{PhysRevResearch.4.043216}. These computer-based approaches are capable of being decoupled from human intuition and consequently have often proven effective at uncovering solution spaces that have never been previously explored. When na\"ive brute force searches are prohibitive, sophisticated algorithms, such as those used in reinforcement learning, have been able to perform at a level that exceeds human capacity, as demonstrated by the strength of computers playing games such as chess and Go~\cite{Silver2016,Silver2017,doi:10.1126/science.aar6404}. The quantum domain is strongly connected with this paradigm due to the analogous exponential complexity underpinning quantum evolution when the system is composed of many constituents~\cite{RL_manyBody,RL_QEC1,RL_QEC2,RL_qCircuit,RL_KapitzaOscillator,Mackeprang2020,RL_Bell}. Examples are finding protocols for quantum communication~\cite{RL_qCommunication}, improving quantum sensors~\cite{RL_qSensor}, and engineering quantum currents~\cite{PhysRevResearch.3.013034}.

As schematically illustrated in Figs.~\ref{fig:schematic}D and 1E, we employ two distinct machine-design approaches: reinforcement learning (RL) and quantum optimal control (QOC), to find lattice position profiles, i.e., control functions, realize desired momentum transformations. In this paper, all control functions are produced by simulating the one-dimensional Schr\"odinger equation for an infinite lattice, with constraints on the lattice depth and total integration time. RL involves the application of an agent which is implemented using a deep neural network. This agent repeatedly proposes a sequence of actions chosen from a predetermined set to discover a sequence that leads to a high terminal reward based on the quantum state fidelity. QOC adopts a different strategy that assumes a Hamiltonian model of the wavefunction evolution and computes the  control sequence that minimizes a given cost function.

The two approaches differ in that RL iterates through many shaking sequences and typically arrives at a different solution in every  run, while for the same control seed  and design parameters, QOC will always return the same optimal control shaking function. It is not particularly useful to compare the two methods per~se, as they operate in different design spaces. In the case of QOC the model is fully specified and the algorithm is deterministic, while in the case of RL, the learning is model free. In model-free learning, it is possible, at least in principle, to perform closed-loop learning on an experimental system whose dynamical equations are completely unknown. Notwithstanding these considerations, both learning methods have proven to be outstanding at finding high quality solutions in our design landscape. Acceptable simulated solutions for beamsplitters and mirrors typically achieve $\geq$97\% fidelity and are often $\geq$99\%.

\begin{figure*}[!ht]
    \centering
    \includegraphics[width=0.75\textwidth]{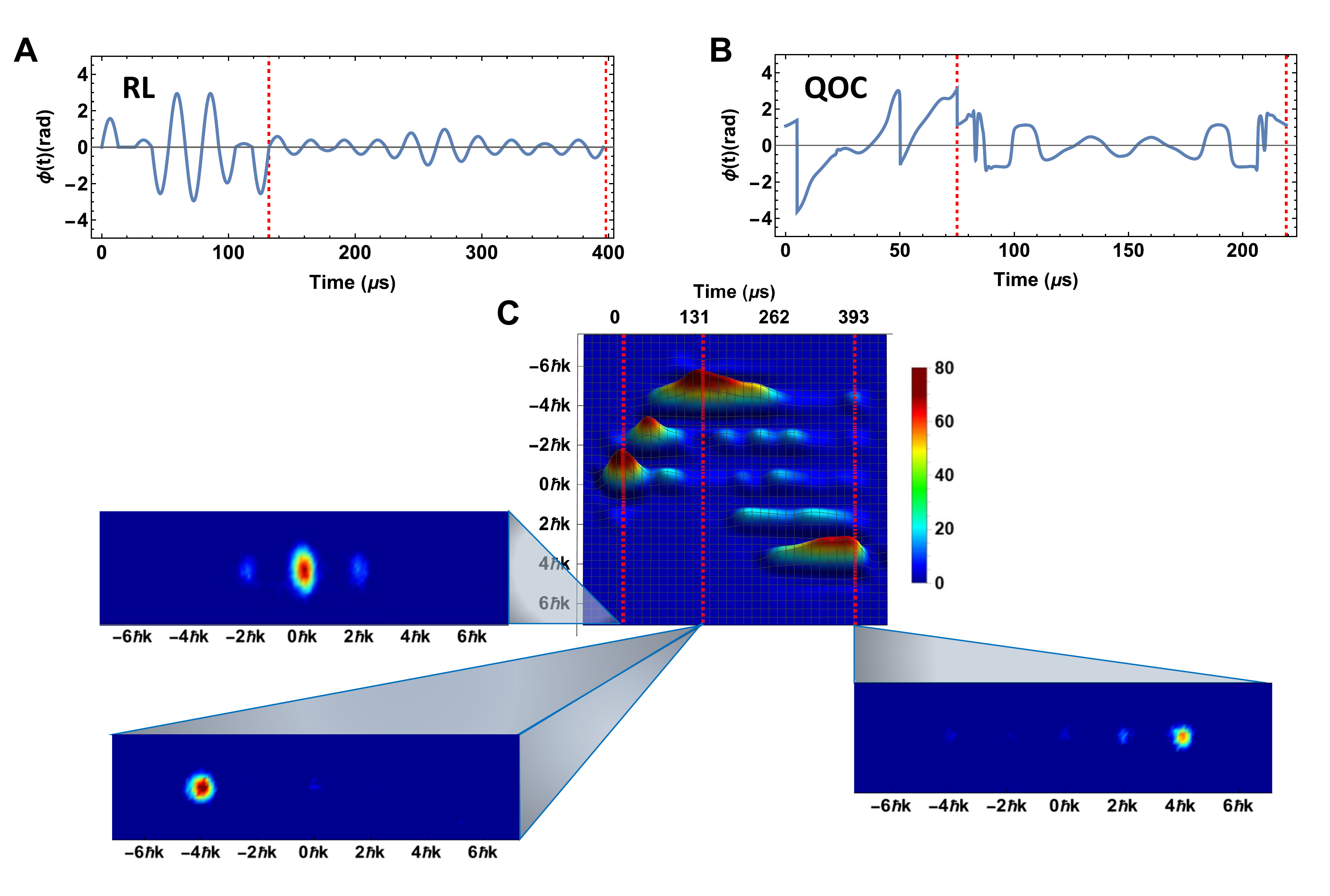}
    \caption{Mirror sequences: Shaking sequences for the 100/0 beamsplitter and mirror for (A) quantum optimal control (QOC) (B) reinforcement learning (RL). Red dashed lines indicate the start and end of the mirror sequence. (C) The cascaded time of flight (TOF) images for RL, as seen in experiment, with pop outs showing the initial Bose-Einstein condensate (BEC) loaded into the lattice, then after the 100/0 beamsplitter, and finally, after the mirror
 is applied. The momentum transfer from $-4\hbar k$ to $4\hbar k$ by the mirror is apparent. Analogous results are observed for sending the atoms along the other path, and/or using QOC control functions.}
    \label{fig:mirror}
\end{figure*}

\vspace*{-1pc}
\subsubsection*{Beamsplitter design}
\vspace*{-.75pc}

We apply both RL and QOC methods to implement offline machine-designed solutions for interferometer components. The first component in the interferometer sequence is an atomic beamsplitter, which places each atom in a superposition of traveling in both directions at the same time. The protocol that we implement was previously developed in Ref~\cite{PhysRevResearch.3.033279} using a 10$E_R$ deep lattice, which is the lattice depth used exclusively for all the experimental sequences presented here. The lattice depth is expressed in terms of the recoil energy, $E_R=\hbar^2 k^2/2m$, where~$m$ is the atomic mass of $^{87}$Rb. With this depth, we realize an approximate beamsplitter by targeting the third excited Bloch state $|n=3, q=0\rangle\equiv|3\rangle$, where $n$ is the band index and $q$ is the quasi-momentum of the periodic lattice. This target state was chosen since it is simultaneously an eigenstate of the lattice and is primarily composed of the two $\pm 4\hbar k$ momentum states with equal weight ($\approx$ 47\% in each). An additional motivation for selecting this target state is that the atoms are excited into the conduction band, enabling them to cover large distances during the transport segment of the interferometer sequence. 

Figures.~\ref{fig:expt_bs}(A)-2(D) show the outcomes following the application of the machine-designed beamsplitter shaking function. Although the shaking functions produced by RL and QOC are dramatically different, the measured fidelities to the target beamsplitter  are both $\geq 95\%$~\cite{PRXQuantum.2.040303}. Both RL and QOC can produce beamsplitters having total shaking times much shorter than than the recoil period, $\tau_R=2\pi/\omega_R\sim 316\,{\rm \mu s}$. We note that these are very short component timescales, of the order of the vibrational period of the atoms in the potential wells of the lattice.

\vspace*{-1pc}
\subsubsection*{Mirror design}
\vspace*{-.75pc}

Designing a shaking function that serves as a mirror is a substantially different problem than the beamsplitter case. For the latter, the design seeks to transform an initial wavefunction into a final target wavefunction.   In the mirror case, the target is a unitary operation~\cite{PhysRevResearch.3.033279} rather than a state. Mathematically, this means a mirror unitary, $U_M$, is designed to perform the following transformation:
\begin{equation}
 U_M \Bigl(\,\alpha |p_0 \rangle  + \beta |{-p_0} \rangle \,\Bigr) \Longrightarrow \beta |p_0 \rangle  +  \alpha |{-p_0} \rangle 
\end{equation}
where $|{\pm p_0}\rangle=\frac1{\sqrt{2}}\bigl(|3\rangle\pm|4\rangle\bigr)$ are the approximately pure $\pm4\hbar k$ momentum eigenstates and channel fidelity~\cite{Pedersen_2007} is used as the metric of performance. Using channel fidelity  restricts the unitary optimization to the subspace spanned by $\{|p_0\rangle,|{-p_0}\rangle\}$, which is the only part of the quantum space that is important for this mirror.

To validate mirror performance we first design a shaking sequence that produces a 100/0 beamsplitter (sending all the atoms to the $|p_0\rangle$ state) or a 0/100 beamsplitter (sending all the atoms to the $|{-p_0}\rangle$ state). This allows us to observe the quality of the momentum reflection directly. Figure.~\ref{fig:mirror}  presents  the experimental evolution from the ground state to a 100/0 beamsplitter state and the subsequent mirror action.

\vspace*{-1pc}
\subsubsection*{The Michelson Interferometer}
\vspace*{-.75pc}

\begin{figure*}[!ht]
    \centering
    \includegraphics[width=0.7\textwidth ]{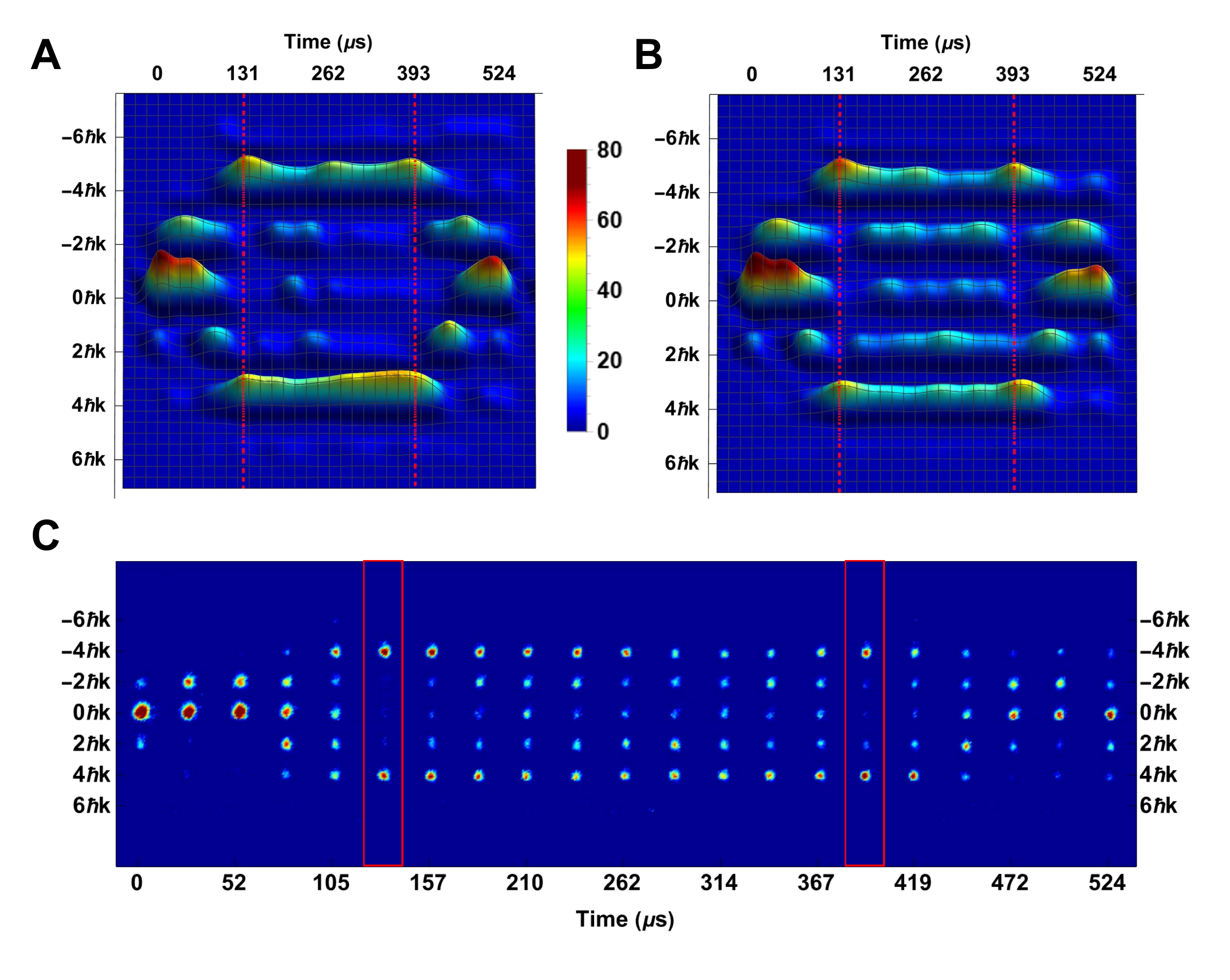}
    \caption{Interferometer sequences: (A) Experimental momentum probability distributions as observed from time of flight (TOF) expansion and cascaded together to give a representation of the interferometer evolution. The color bar is expressed in units of normalized population percentage and red lines indicate the boundaries between the beamsplitter, mirror, and recombiner sequences, which are sandwiched together with no propagation delays. The design problem specifies the desired wavefunction only on these boundary lines. (B) Same depiction as (A) but for the theoretical quantum evolution. (C) Raw experimental TOF images stitched together. Note that the observed asymmetry between positive and negative momenta is anticipated since shaking solutions immediately break the inversion symmetry by initially shifting the lattice in one direction.}
    \label{fig:expt_Michelson}
\end{figure*}

Equipped with beamsplitter and mirror protocols, a full interferometer sequence is stitched together, effectively splitting, propagating, mirroring, propagating, and finally recombining for phase readout, as shown in Figs.~\ref{fig:expt_Michelson}(A)-4(C). We impose time reversal symmetry so that only the first half of the interferometer sequence need be designed.

Following the entire interferometer sequence, the atoms are largely returned to the ground state of the lattice. This is anticipated if the components are implemented accurately since, ideally, in the absence of an externally applied acceleration signal, the initial  state should be perfectly recovered  at the end of the sequence. The fact that we observe this, as shown in Fig.~\ref{fig:expt_Michelson}, indicates not only that the devices are operating very close to what was intended in terms of both phase and amplitude wave function evolution but also that any effects not modeled by the design equations do not adversely affect solutions on these time scales.

\vspace*{-1pc}
\subsubsection*{Interferometer response to acceleration}
\vspace*{-.75pc}

\begin{figure*}[!htb]
    \centering
    \includegraphics[width=1\textwidth]{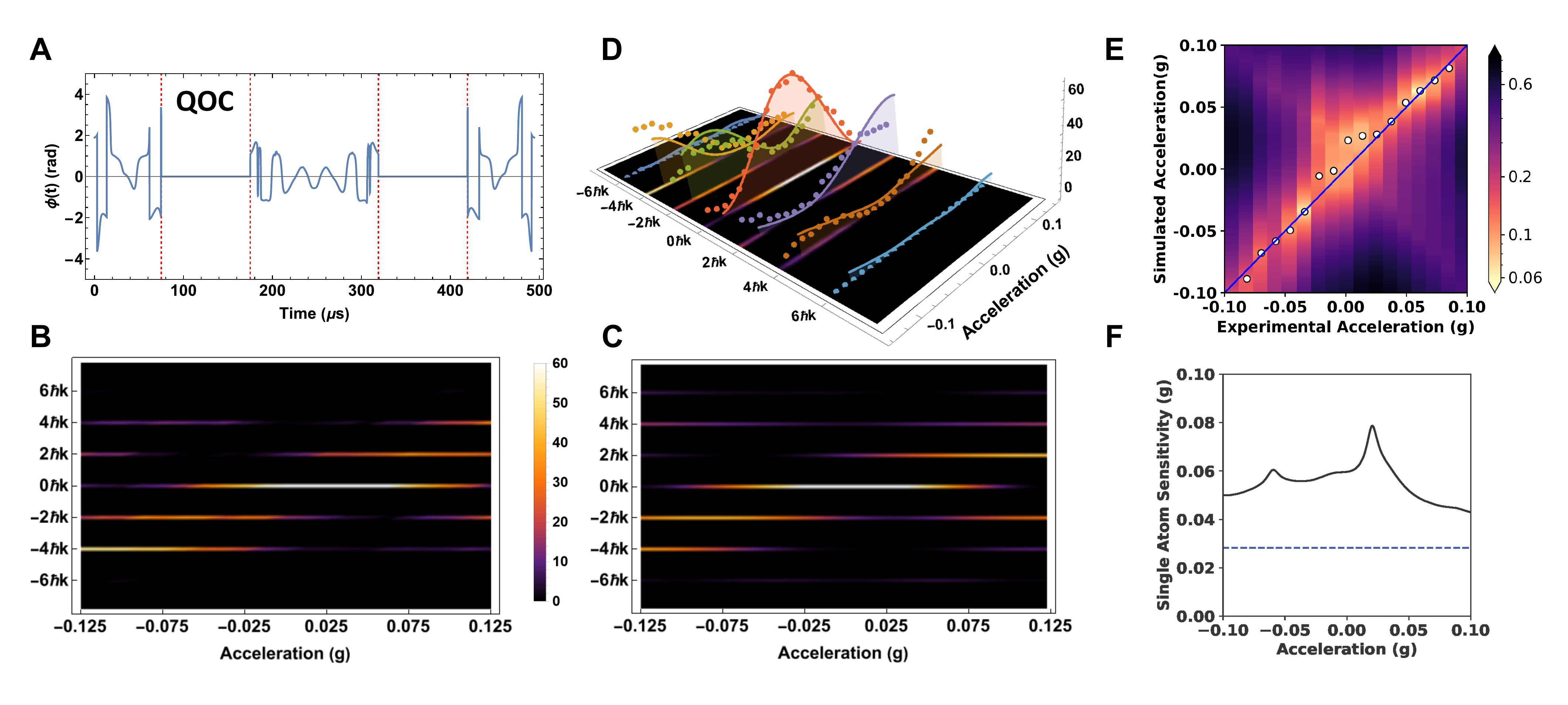}
    \caption{Accelerometer measurements: (A) The quantum optimal control (QOC) interferometer control sequence, where red lines separate interferometer components. Full interferometer sequences were conducted over a set of equally spaced acceleration values from -0.125$g$ to +0.125$g$, and time of flight (TOF) momentum populations were measured and stitched together to generate the contoured scan depicted in (B). The expected result from the theory calculation is shown in (C). Fringes for each momentum component are depicted in (D). The experimental data points (dots) compare well with the anticipated theory (solid lines). The theory lines shown are not fits to the data and contain no free parameters with the exception of the overall shift in the zero acceleration. Analysis of bias offset and sensitivity: (E)~Jensen-Shannon divergence shown on a logarithmic scale. Low values $\simeq 0$ correspond to almost perfect coincidence of the measured distribution with theory (light regions), and high values $\simeq 1$ correspond to poor coincidence (dark regions). The minimum at each recorded data distribution is the maximum likelihood estimator for acceleration (white circles). (F)~Single-atom sensitivity (smaller values are more sensitive) from theory for this shaking sequence compared with the ideal (instantaneous component and perfect momentum splitting) limit with the same total time (blue dashed line).}
    \label{fig:scan}
\end{figure*}

In a standard interferometer, the output is dependent upon an applied acceleration. The sensitivity to this acceleration is proportional to the enclosed space-time area. The shaken lattice approach allows us to adjust the enclosed area by incorporating transport stages between the interferometer components, see Fig.~\ref{fig:schematic}(A) During the transport stages the lattice position is held fixed in space.  Here we allow the atoms to propagate for 100~$\mu$s in each direction. We keep this time relatively short to ensure that any potential effects of transverse diffusion of the atoms, walk-off from the CDT intersection, and atom-atom interactions, will not significantly impact the sensor performance. An acceleration is imposed onto the atoms via a frequency chirp applied to one of the lattice AOMs.  Although both design methods produce comparable results, we display only the QOC results for the sake of brevity, with the control function shown in Fig.~\ref{fig:scan}(A). The resulting momentum state interference fringes as a function of acceleration are shown in Fig.~\ref{fig:scan}(B).
The experimental momentum populations measured over the applied acceleration range compare well with the theory presented in Fig.~\ref{fig:scan}(C). Note that the only fitting parameter is a small overall shift of the zero value in the applied accelerations which is accounted for in the data presented in Figs.~\ref{fig:scan}(D) and 5(E). This shift is caused by a slight tilt ($<1^\circ$) in our lattice beams, which induces an acceleration offset.
\newline
\indent There are several significant aspects of the resulting fringes, as illustrated in Fig.~\ref{fig:scan}(D), that we draw attention to. The output of our accelerometer is the momentum state population fractions $p\in\{-6\hbar k, -4\hbar k, \ldots, 6\hbar k\}$ viewed after recombination and TOF, rather than the interference signal of two recombined clouds as in conventional Bragg interferometry. The fraction of the total population in these seven momentum states, $P(p|a)$, provides a unique fingerprint for each acceleration $a$, which allows the value of the acceleration to be determined from the measured momentum populations using statistical methods. The populations are not symmetric about the applied accelerations, enabling the discrimination of the direction of the applied signal. The fringe periodicity, which determines the dynamic range of the device, can easily be tuned by simply changing the transport time, with longer times giving higher sensitivity and smaller bandwidth, and shorter times giving lower sensitivity and higher bandwidth. This ability to modify the system performance metrics at will is particularly attractive for applications in which the measurement scenario is dynamic. 

\vspace*{-1pc}
\subsubsection*{Sensitivity}
\vspace*{-.75pc}
The performance of the atom accelerometer can be characterized by the the Jensen-Shannon (JS) divergence  illustrated in Fig.~\ref{fig:scan}(E). The JS divergence quantifies  the degree of consistency between the simulation and the experimental measurements~\cite{lin_OriginalJSDivergence}.  
An unbiased estimator for the measured acceleration  is the maximum likelihood, which corresponds to the minimum of the JS divergence~\cite{csiszar2003information} at each measured data point and is shown as white circles. The degree to which the calibrated points lie in proximity to the diagonal line is a measure of sensitivity. To quantify this, we compute the single atom sensitivity $1/\sqrt{I(a)}$, where $I(a)$ is the classical Fisher information,
\begin{equation}
    I(a)=\sum_{p\in\{-6\hbar k,\ldots,6\hbar k\}}\frac1{P(p|a)}\left[\frac{\partial P(p|a)}{\partial a}\right]^2   
\end{equation}
which is shown in Fig.~\ref{fig:scan}(F). For reference, we bound this by the ideal single atom sensitivity that would result if the beamsplitter and mirror components took zero time and the device operated in free space with exactly $\pm4\hbar k$ splitting, thereby giving a perfect two-arm interferometer.

\vspace*{-1pc}
\subsection*{Discussion And Conclusions}
\vspace*{-.75pc}
In this paper we have demonstrated the application of advanced machine-design methods to find a class of quantum design solutions for phase modulation, that is, the shaking of an optical lattice. We have shown the ability to produce a variety of beamsplitters (50/50, 100/0, and 0/100), mirrors, and recombiners all of which have been realized experimentally. When the designed components were cascaded, the resulting interferometer protocols were analyzed through the application of a lattice acceleration and the subsequent measurement of the output diffraction pattern.

The interferometer demonstrated here has a  round-trip time of $\leq$ 500~$\mu$s. The transport time in the lattice (100~$\mu$s) implies that the atoms only move on the order of $2.5\,\mu$m. This is important because the sensitivity of an accelerometer scales with the area enclosed in space-time. If the transport time were increased from  100~$\mu$s to, say, 0.1~s, the atomic wavepackets would be displaced by $\approx$2.5 mm;  this would increase sensitivity by a factor of order $(10^3)^2$. According to Fig.~\ref{fig:scan}(F), this implies a possible single-atom sensitivity of $\approx6\times10^{-8}g$. This value is not a sensitivity limit, since each fringe can be more precisely determined by statistical averaging using multiple atoms, multiple passes, and multiple trials~\cite{kim2022second}. The anticipated performance at the shot noise limit can, in general, be computed from the single-atom sensitivity by dividing by a factor of $\sqrt{N}$ for $N$ independent measurements. Also note that what we present here is a classical limit, so that entangling and squeezing the atomic quantum state could potentially improve the sensitivity further. 

It should also be noted that because we use machine learning for design, protocols can be constructed to mitigate the adverse effects of laser noise and atom interactions~\cite{chihthesis}. This can be done by including each of these effects in the reward function used during learning, a process that involves application of the multiparameter quantum Fisher information matrix~\cite{Reilly_2023}. This leads to design protocols that are extremely sensitive to acceleration, while being simultaneously insensitive to imperfections, such as lattice depth and/or atom number fluctuations, commonly classified as nuisance parameters. The ability to design around imperfections is a feature that can be used to make optical-lattice interferometry attractive for deployment in environments where robustness is needed.

In summary, in this paper, we have realized one of a class of atom interferometers that are real-time reconfigurable, have a  compact form factor, and the potential to extend the application domain of atom interferometry. For more details, see the Supplemental Material. 

\vspace*{-1pc}
\subsubsection*{Acknowledgements}
\vspace*{-.75pc}

We would like to thank Liang-Ying Chih, Malcolm Boshier, Ceren Uzun, Katarzyna Krzyżanowska, and Mantas Naris for many helpful discussions. This paper was supported by
NSF OMA Grant No. 1936303; NSF PHY Grant No. PHY 2317149; NSF OMA Grant No. 2016244; and NSF PHY Grant No. 2207963, and by a Sponsored Research Agreement between the University of Colorado Boulder and Infleqtion. Dana Z. Anderson retains stock in, and serves on the Board of Directors of, ColdQuanta Inc., dba Infleqtion

%\bibliography{ref}

\end{document}

\section*{Supplementary Material}

\vspace*{-1pc}
\subsubsection*{BEC production}
\vspace*{-.75pc}
Our experimental system is constructed around a ColdQuanta physics station which houses a custom double magneto-optical trap (MOT) cell. Because we work with $^{87}$Rb, a 780 nm laser system allows us to initially cool and trap from background vapor emitted from a rubidium dispenser. We use a series of Vescent distributed Bragg reflector (DBR) lasers, with a master laser locked to absorption spectroscopy, and two additional lasers that are offset locked from the master to the desired repump and cooling transitions. A push beam loads cooled atoms from the lower 2D MOT glass chamber into an upper 3D MOT science cell, where, in a matter of seconds we routinely produce MOT's consisting of over $10^9$ atoms. Following sufficient MOT load, we implement a compressed MOT (CMOT) stage followed by polarization gradient cooling (PGC), both of which involve further shifting the laser detuning and magnetic field gradients. At the end of PGC, we have around $6\times 10^8$ atoms at about 20~$\mu$K. Further cooling is then performed in an all-optical approach, which involves initially loading atoms into crossed dipole trap (CDT). An IPG 1064nm 30 W laser is used to construct the CDT which consists of two beams which intersect at $40^\circ$ and have 55~$\mu$m  waists. The second dipole beam is generated by re-passing the first beam through the science cell, so there is no independent power control between the two beams. The CDT is turned on during the CMOT and PGC stages. During the last millisecond of PGC, the repump light is extinguished, thereby pumping atoms into the $F=1$ state. This means that to image, we must flash repump light back on which transfers the atoms back into the $F=2$ imaging state. After PGC and a 50 ms hold, we then evaporatively cool the atoms by linearly ramping down the CDT power in three successive stages. In this way, the CDT power is adiabatically ramped from about 15 W down to 100 mW. Our total evaporation sequence takes approximately 12 s, after which we typically produce BECs with as many as $10^5$ atoms with temperatures below 20 nK. Before performing interferometer experiments we intentionally reduce the atom number to approximately $2 \times 10^4$ in order to simplify the theory comparison. Due to the all-optical evaporation approach, the atoms are in a mixture of magnetic hyperfine sublevels. Note that it is typically the case in precision metrology for inertial sensing that one needs to pay careful attention to stray magnetic fields and to migitate these effects it is preferable to be in the m$_{F}$=0 state. This is typically achieved by either optical pumping or by applying a magnetic field gradient during the last stage of evaporation.

\vspace*{-1pc}
\subsubsection*{1D lattice}
\vspace*{-.75pc}
Following evaporation, our 1D lattice is adiabatically ramped on in 800 $\mu$s and atoms are loaded into the Bloch ground state of the lattice. The lattice itself is constructed from a Vescent 852nm DBR laser that is locked to spectroscopy and is further amplified via a NewFocus VAMP TA-7600. The lattice beams are co-linear to the first pass of the CDT and currently there is no active intensity stabilization.  We apply phase modulation to the optical lattice by updating the RF frequency driving one of the lattice AOMs. We utilize an Agilent 33622A arbitrary waveform generator and typically sample the modulation waveforms of a shaking protocol with 50 ns resolution, which we have found to be sufficient to realize high experimental fidelities. When there is no acceleration applied, the second AOM is driven from the same waveform generator, but with a constant 80~MHz carrier frequency during implementation of shaking sequences. To apply an acceleration signal to the atoms, the second AOM is driven with a linear frequency sweep away from its carrier. The frequency sweep and the start of the shaking sequence for the atom interferometer are both simultaneously triggered following the initial lattice load. Although there is no active phase stabilization of the optical lattice, we do validate the effect of the shaking sequences on the lattice directly via an optical Michelson interferometer upstream in the optical path. 
\vspace*{-1pc}
\subsubsection*{Reinforcement learning}
\vspace*{-.75pc}
The full details of the RL approach are presented in Ref.~\cite{PhysRevResearch.3.033279}. This paper outlines most of the protocols that have been implemented here, including preparation in the ground state of a $10E_R$ deep lattice, and the process for determining machine-trained beamsplitter and mirror solutions. This work also evaluated the sensitivity through the calculation of Fisher information and Bayesian reconstruction. 

The learning protocol for RL is based on solving the Bellman optimality equation~\cite{Bellman57}. Our RL algorithm is Double DQN~\cite{DQNNature,doubleDQN}, using temporal differencing to learn the action-value function associated with a given state, with an epsilon greedy policy guiding the agent decisions, and with an update to two neural networks (Q-network and target network) to produce the Q-value estimates. 

There are a few essential differences from the previous work. In particular, the learned lattice phase, $\phi(t)$, was sinusoidal for all our shaking solutions given here, with the amplitude allowed to take on any one of a discrete set of 32 values, i.e., $\{\pi j/32:j=0,\ldots,31\}$, at each half cycle.  The frequency of the sinusoid in all cases was approximately $12\,\omega_R$, which was found empirically to provide high fidelity learned sequences. This makes sense since it corresponds to the energy gap  between the kinetic energy of a free particle with $2\hbar k$ and  with~$4\hbar k$. The second half of the mirror and the recombiner were generated by applying time reversal. Furthermore the channel fidelity was computed within the subspace of  $\ket{3}$ and~$\ket{4}$, a slight change from the momentum subspace used previously. Finally, we use the norm of the even and odd parity components of the momentum distribution as the feature of the state for the neural network input layer.

\vspace*{-1pc}
\subsubsection*{Quantum optimal control}
\vspace*{-.75pc}

The QOC problem can be written as
\begin{equation}\label{eq:qoc}
\text{minimize} \, \mathcal J\bigl(\ket{\psi(T)}\bigr)+\, \int_0^T \ell\bigl(\phi(t)\bigr)dt
\end{equation}
where $\ket{\psi(t)}$ is the state which evolves according to the Schr\"odinger evolution
\begin{equation}\label{eq:lattice}
    i\hbar \frac{\partial\ket{\psi(t)}}{\partial t}=\left[\frac{\hat{p}^2}{2m}+V_L\cos
    \bigl(2k\hat{x}+\phi(t)\bigr)\right]\ket{\psi(t)},
\end{equation}
and $\hat x$ and $\hat p$ denote the canonical position and momentum operators. The goal of QOC is to find the optimal shaking function $\phi(t)$ so that the state $\ket{\psi}$ starts at an initial state $\ket{\psi_0}$ and evolves to a target state $\ket{\psi_T}$. This is solved numerically by the Q-PRONTO algorithm, as detailed in Ref.~\cite{PhysRevA.105.032605}.

The cost function in Eq.~(\ref{eq:qoc}) consists of two parts; an incremental cost $\ell\bigl(\phi(t)\bigr)$, and a terminal cost $\mathcal J\bigl(\ket{\psi(T)}\bigr)$, where $T$ is the time horizon. The terminal cost function for the beamsplitter is
\begin{equation}
    \mathcal   J\bigl(\ket{\psi(T)}\bigr) = 1 - |\bra{\psi(T)}3\rangle|^2
\end{equation}
%\begin{equation}
 %   \mathcal   J\bigl(\ket{\psi(T)}\bigr)=\bra{\psi(T)}P\ket{\psi(T)},
%\end{equation}
i.e., the Hilbert-Schmidt distance between the terminal state $\ket{\psi(T)}$ and the target state $\ket{3}$. The incremental cost is defined as
\begin{equation}\label{eq:control_effort}
\ell\bigl(\phi(t)\bigr)=\frac{r}{2}\phi^2(t),
\end{equation}
where $r > 0$ is an adjustable parameter that penalizes excessive utilization of the control input (shaking function). For the mirror, a similar procedure follows, except the cost function is computed within the operational channel.

\begin{figure*}[!t]
    \centering
    \includegraphics[width=0.8\textwidth]{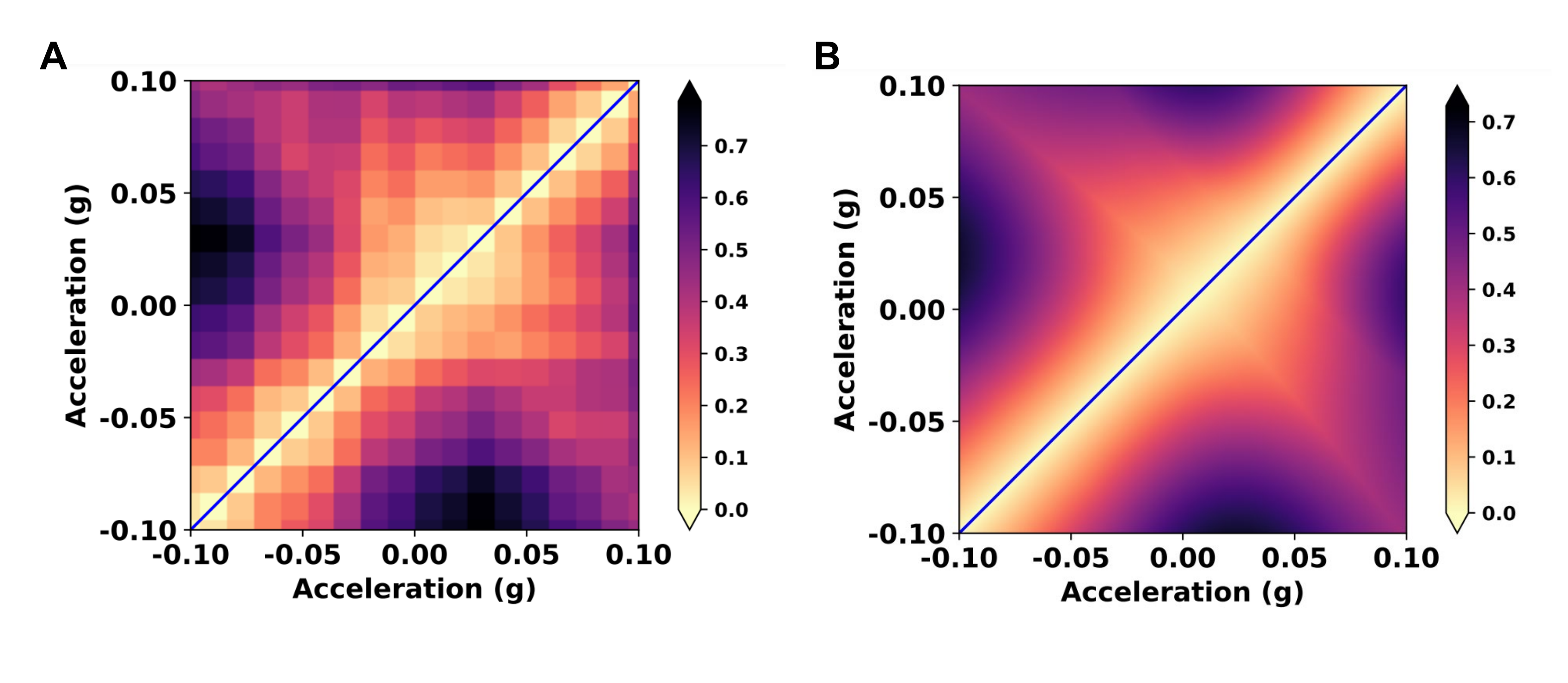}
    \caption{(A) The Jensen-Shannon divergence for the experimental autocorrelation. This representation shows the degree to which the momentum distribution is unrepeated for different accelerations and therefore indicates the ability to uniquely specify the acceleration parameter from a single measurement. The analogous autocorrelation for the theoretical autocorrelation is shown in (B). For both theory and experiment, the low values on the diagonal are always anticipated, but on the anti-diagonal are a consequence of the symmetry of the problem with respect to a spatial inversion.}
    \label{fig:autocorrelation}
\end{figure*}

\vspace*{-1pc}
\subsubsection*{Connecting simulated accelerations to gravity}
\vspace*{-.75pc}
As discussed in the main text, we analyze the accelerometer performance by applying a differential frequency ramp to the optical lattice lasers. This is equivalent to an acceleration of the inertial sensor, and can therefore be related to gravitational free-fall via the equivalance principle. To see this, we start with the Hamiltonian for an atom in an optical lattice in the presence of a gravitational potential
\begin{equation}
    \hat{H}^{(1)}(t) = \frac{\hat{p}^2}{2m}+mg\hat{x}+V_L\cos\bigl(2k\hat{x}+\phi(t)\bigr)
\end{equation}
where the term $mg\hat x$ arises from a gravitational acceleration~$g$. We transform into a rotating frame by separating the Hamiltonian into two parts, $\hat H^{(1)}(t)=\hat H_0^{(1)} + \hat V^{(1)}(t)$
with $\hat H_0^{(1)}=mg\hat x$ and $V^{(1)}(t) = \hat{p}^2/2m+V_L\cos\bigl(2k\hat{x}+\phi(t)\bigr)$. In this rotating frame, the Hamiltonian is given by
\begin{equation}
    \hat{H}^{(2)}(t) = \frac{{\hat{\pi}(t)}^2}{2m}+V_L\cos\bigl(2k\hat{x}+\phi(t)\bigr),\qquad
\end{equation}
where $\hat{\pi}(t)\equiv\hat{p}-mgt$ is the kinematic momentum. 

We can continue with one more rotating frame transformation to move away from the reference frame of the lattice. We separate the Hamiltonian again, $\hat H^{(2)}(t) = H_0^{(2)}(t) + V^{(2)}(t)$ with $\hat H_0^{(2)}(t) = \bigl({\hat{\pi}(t)}^2-\hat{p}^2)/2m$ and $V^{(2)}(t) = \hat{p}^2/2m+V_L\cos\bigl(2k\hat{x}+\phi(t)\bigr)$. In the new rotating frame, the Hamiltonian is given by
\begin{equation}
\hat{H}^{(3)}(t) = \frac{\hat{p}^2}{2m}+V_L\cos\bigl(2k\hat{x}+\beta(t)\bigr),
\end{equation}
where $\beta(t)=\phi(t)+2kx_g$ for $x_g=-\frac12 gt^2$. This picture can be interpreted as a frame falling with the atoms, and therefore the lattice appears to accelerate in the opposite direction according to constant acceleration kinematics. The equivalence of these three frames enables one to evaluate the performance of the sensor as a gravimeter by appropriate ramp of the lattice phase.

\vspace*{-1pc}
\subsubsection*{Maximum likelihood estimators}
\vspace*{-.75pc}

The relative populations of each momentum in the sample space $p\in\{-6\hbar k,\ldots,6\hbar k\}$ represent probability distributions, and therefore, one can search for the theory that most closely matches the experimentally determined distribution. There are many notions of closeness in probability theory, but the most common are called divergences, which measure logarithmic differences often relating to entropy. The relative entropy is one example of a divergence measure and is often called the Kullback-Leibler (KL) divergence, $D_{\rm{KL}}(P\Vert Q) = \sum_p P(p) \log_2\bigl(P(p)/Q(p) \bigr)$, for two distributions $P$ and $Q$. KL divergence has the downside of being asymmetric and unbounded, and therefore, to be defined requires $P(p)=0$ if and only if $Q(p)=0$, which is not always guaranteed within experimental data.

An alternative is the Jensen-Shannon (JS) divergence, 
\begin{equation}
    D_{\rm{JS}}(P\Vert Q)=\frac12\bigl(D_{\rm{KL}}(P\Vert M)+D_{\rm{KL}}(Q\Vert M)\bigr)
\end{equation} 
defined by the mixture distribution $M(p)=\bigl(P(p)+Q(p)\bigr)/2$, and is bounded by the interval $[0,1]$. The JS divergence provides a smoothed and symmetrized version of the KL divergence~\cite{JSDiv_HilbertSpace}, and can be understood as the mutual information between a binary indicator of having measured from $P$ or $Q$, and the measured value itself. If $P$ and $Q$ are identical, then it doesn't make a difference if one samples from $P$ or $Q$, and the JS divergence is zero. Alternatively, if the probability distributions are entirely incommensurate, the JS divergence goes to one as the indicator perfectly discriminates. In Fig.~5E, for each observed distribution $P_\text{exp}(p)$, we searched the full space of theory distributions, $P_\text{thy}(p|g)$, for the value of $g$ that leads to a minimal value of the JS divergence, i.e., $\argmin_g\bigl(D_{\rm{JS}}(P_\text{exp}(p)\Vert P_\text{thy}(p|g))\bigr)$. This calculates an information-projection~\cite{nielsen2018information,khanna2015sparse,khanna2017information}, which allows the inferred experimental distribution to be mapped to the closest theory distribution.

As shown in Fig.~\ref{fig:autocorrelation} this method can also be used to indicate the degree to which each measured value of the acceleration gives rise to a unique distribution. One of the most important features is the ability to recognize aliasing, where the repetitions due to the recurrence of diffraction patterns in quantum interference can limit the dynamic range of the instrument. In our case, it is apparent that the inversion symmetry leads to small values of the JS divergence close to the off diagonal for both theoretical and experimental autocorrelations, as expected, but the maximum likelihood solution, shown in Fig.~5E, always lies in close proximity to the diagonal.

\newpage

\clearpage